\pdfoutput=1
\documentclass[12pt,a4paper]{article}

\usepackage{ifthen} 
\newboolean{pdflatex}
\setboolean{pdflatex}{true} 

\newboolean{articletitles}
\setboolean{articletitles}{true} 

\newboolean{uprightparticles}
\setboolean{uprightparticles}{false} 

\def\paperauthors{LHCb collaboration} 
\def\paperasciititle{Search for CP violation in the phase space of D0 -> pi-pi+pi0 decays with the energy test} 
\def\papertitle{Search for \CP violation in the phase space of $\Dz \to \pim\pip\piz$ decays with the energy test} 
\def\paperkeywords{{High Energy Physics}, {LHCb}} 
\def\papercopyright{\the\year\ CERN for the benefit of the LHCb collaboration} 
\def\paperlicence{CC BY 4.0 licence}
\def\paperlicenceurl{https://creativecommons.org/licenses/by/4.0/}
\def\tsymbol {\ensuremath{T}\xspace}
\def\tvalue {\ensuremath{T}-value\xspace}

\usepackage[top=1in, bottom=1.25in, left=1in, right=1in]{geometry}

\usepackage{microtype}
\usepackage{lineno}  
\usepackage{xspace} 
\usepackage{caption} 

\usepackage{graphicx}  
\usepackage{color}
\usepackage{colortbl}
\graphicspath{{./figs/}} 

\usepackage{amsmath} 
\usepackage{amssymb}
\usepackage{amsfonts}
\usepackage{gensymb}
\usepackage{upgreek} 

\newcommand*\patchAmsMathEnvironmentForLineno[1]{%
\expandafter\let\csname old#1\expandafter\endcsname\csname #1\endcsname
\expandafter\let\csname oldend#1\expandafter\endcsname\csname
end#1\endcsname
 \renewenvironment{#1}%
   {\linenomath\csname old#1\endcsname}%
   {\csname oldend#1\endcsname\endlinenomath}%
}
\newcommand*\patchBothAmsMathEnvironmentsForLineno[1]{%
  \patchAmsMathEnvironmentForLineno{#1}%
  \patchAmsMathEnvironmentForLineno{#1*}%
}
\AtBeginDocument{%
\patchBothAmsMathEnvironmentsForLineno{equation}%
\patchBothAmsMathEnvironmentsForLineno{align}%
\patchBothAmsMathEnvironmentsForLineno{flalign}%
\patchBothAmsMathEnvironmentsForLineno{alignat}%
\patchBothAmsMathEnvironmentsForLineno{gather}%
\patchBothAmsMathEnvironmentsForLineno{multline}%
\patchBothAmsMathEnvironmentsForLineno{eqnarray}%
}

\usepackage{hyperxmp}

\usepackage[pdftex,
            pdfauthor={\paperauthors},
            pdftitle={\paperasciititle},
            pdfkeywords={\paperkeywords},
            pdfcopyright={Copyright (C) \papercopyright},
            pdflicenseurl={\paperlicenceurl}]{hyperref}

\usepackage[colorinlistoftodos,textsize=scriptsize]{todonotes}

\usepackage[bottom,flushmargin,hang,multiple]{footmisc}

\usepackage[all]{hypcap} 

\usepackage{xspace} 
\usepackage{upgreek}

\def\lhcb   {\mbox{LHCb}\xspace}

\def\MagUp {\mbox{\em Mag\kern -0.05em Up}\xspace}

\ifthenelse{\boolean{uprightparticles}}%
{

 \def\Ppi         {\ensuremath{\uppi}\xspace}

 \def\PDelta      {\ensuremath{\Delta}\xspace}                 
 \def\PXi         {\ensuremath{\Xi}\xspace}                 
 \def\PLambda     {\ensuremath{\Lambda}\xspace}                 
 \def\PSigma      {\ensuremath{\Sigma}\xspace}                 
 \def\POmega      {\ensuremath{\Omega}\xspace}                 
 \def\PUpsilon    {\ensuremath{\Upsilon}\xspace}
 \let\oldPi\Pi
 \def\PPi         {\ensuremath{\oldPi}\xspace}

 \def\PB      {\ensuremath{\mathrm{B}}\xspace}                 
                  
 \def\PD      {\ensuremath{\mathrm{D}}\xspace}

 \def\PK      {\ensuremath{\mathrm{K}}\xspace}

 \def\Pb      {\ensuremath{\mathrm{b}}\xspace}                 
 \def\Pc      {\ensuremath{\mathrm{c}}\xspace}

 \def\Pi      {\ensuremath{\mathrm{i}}\xspace}

 \def\Ps      {\ensuremath{\mathrm{s}}\xspace}

 \def\thebaroffset{0.0em}
}
{

 \def\Ppi         {\ensuremath{\pi}\xspace}

 \mathchardef\PDelta="7101
 \mathchardef\PXi="7104
 \mathchardef\PLambda="7103
 \mathchardef\PSigma="7106
 \mathchardef\POmega="710A
 \mathchardef\PUpsilon="7107
 \mathchardef\PPi="7105
                  
 \def\PB      {\ensuremath{B}\xspace}                 
                  
 \def\PD      {\ensuremath{D}\xspace}

 \def\PK      {\ensuremath{K}\xspace}

 \def\Pb      {\ensuremath{b}\xspace}                 
 \def\Pc      {\ensuremath{c}\xspace}

 \def\Pi      {\ensuremath{i}\xspace}

 \def\Ps      {\ensuremath{s}\xspace}

 \def\thebaroffset{0.18em}
}
\newcommand{\offsetoverline}[2][\thebaroffset]{\kern #1\overline{\kern -#1 #2}}%

\makeatletter
\ifcase \@ptsize \relax
  \newcommand{\miniscule}{\@setfontsize\miniscule{4}{5}}
\or
  \newcommand{\miniscule}{\@setfontsize\miniscule{5}{6}}
\or
  \newcommand{\miniscule}{\@setfontsize\miniscule{5}{6}}
\fi
\makeatother

\DeclareRobustCommand{\optbar}[1]{\shortstack{{\miniscule (\rule[.5ex]{1.25em}{.18mm})}
  \\ [-.7ex] $#1$}}

\def\squark    {{\ensuremath{\Ps}}\xspace}

\def\cquark    {{\ensuremath{\Pc}}\xspace}

\def\bquark    {{\ensuremath{\Pb}}\xspace}

\def\pion   {{\ensuremath{\Ppi}}\xspace}
\def\piz    {{\ensuremath{\pion^0}}\xspace}
\def\pip    {{\ensuremath{\pion^+}}\xspace}
\def\pim    {{\ensuremath{\pion^-}}\xspace}

\def\kaon    {{\ensuremath{\PK}}\xspace}

\def\KorKbar {\kern \thebaroffset\optbar{\kern -\thebaroffset \PK}{}\xspace}

\def\Kp      {{\ensuremath{\kaon^+}}\xspace}
\def\Km      {{\ensuremath{\kaon^-}}\xspace}

\def\KS      {{\ensuremath{\kaon^0_{\mathrm{S}}}}\xspace}

\def\Dbar    {{\ensuremath{\offsetoverline{\PD}}}\xspace}
\def\D       {{\ensuremath{\PD}}\xspace}

\def\DorDbar {\kern \thebaroffset\optbar{\kern -\thebaroffset \PD}\xspace}
\def\Dz      {{\ensuremath{\D^0}}\xspace}
\def\Dzb     {{\ensuremath{\Dbar{}^0}}\xspace}
\def\Dp      {{\ensuremath{\D^+}}\xspace}
\def\Dm      {{\ensuremath{\D^-}}\xspace}

\def\DpDm    {\ensuremath{\Dp {\kern -0.16em \Dm}}\xspace}

\def\Dstarp  {{\ensuremath{\D^{*+}}}\xspace}

\def\B       {{\ensuremath{\PB}}\xspace}

\def\BorBbar {\kern \thebaroffset\optbar{\kern -\thebaroffset \PB}\xspace}

\def\Bd      {{\ensuremath{\B^0}}\xspace}

\def\BdorBdbar {\kern \thebaroffset\optbar{\kern -\thebaroffset \Bd}\xspace}

\def\Bs      {{\ensuremath{\B^0_\squark}}\xspace}

\def\BsorBsbar {\kern \thebaroffset\optbar{\kern -\thebaroffset \Bs}\xspace}

\def\Y#1S{\ensuremath{\PUpsilon{(#1S)}}\xspace}

\def\LorLbar     {\kern \thebaroffset\optbar{\kern -\thebaroffset \PLambda}\xspace}

\newcommand{\decay}[2]{\ensuremath{#1\!\to #2}\xspace} 
\def\ra                 {\ensuremath{\rightarrow}\xspace}
\def\to                 {\ensuremath{\rightarrow}\xspace}

\def\order   {{\ensuremath{\mathcal{O}}}\xspace}

\def\CP                {{\ensuremath{C\!P}}\xspace}

\def\AT#1     {\ensuremath{A_{\mathrm{T}}^{#1}}\xspace}           

\def\C#1      {\ensuremath{\mathcal{C}_{#1}}\xspace}                       
\def\Cp#1     {\ensuremath{\mathcal{C}_{#1}^{'}}\xspace}                    
\def\Ceff#1   {\ensuremath{\mathcal{C}_{#1}^{\mathrm{(eff)}}}\xspace}        
\def\Cpeff#1  {\ensuremath{\mathcal{C}_{#1}^{'\mathrm{(eff)}}}\xspace}       
\def\Ope#1    {\ensuremath{\mathcal{O}_{#1}}\xspace}                       
\def\Opep#1   {\ensuremath{\mathcal{O}_{#1}^{'}}\xspace}                    

\newcommand{\nospaceunit}[1]{\ensuremath{\text{#1}}}       
\newcommand{\aunit}[1]{\ensuremath{\text{\,#1}}}       

\newcommand{\tev}{\aunit{Te\kern -0.1em V}\xspace}
\newcommand{\gev}{\aunit{Ge\kern -0.1em V}\xspace}
\newcommand{\mev}{\aunit{Me\kern -0.1em V}\xspace}
\newcommand{\kev}{\aunit{ke\kern -0.1em V}\xspace}
\newcommand{\ev}{\aunit{e\kern -0.1em V}\xspace}
 
\newcommand{\mevc}{\ensuremath{\aunit{Me\kern -0.1em V\!/}c}\xspace}
\newcommand{\gevc}{\ensuremath{\aunit{Ge\kern -0.1em V\!/}c}\xspace}
\newcommand{\mevcc}{\ensuremath{\aunit{Me\kern -0.1em V\!/}c^2}\xspace}
\newcommand{\gevcc}{\ensuremath{\aunit{Ge\kern -0.1em V\!/}c^2}\xspace}

\def\mum  {\ensuremath{\,\upmu\nospaceunit{m}}\xspace}

\def\fb   {\ensuremath{\aunit{fb}}\xspace}
\def\invfb   {\ensuremath{\fb^{-1}}\xspace}

\def\order{{\ensuremath{\mathcal{O}}}\xspace}

\def\gsim{{~\raise.15em\hbox{$>$}\kern-.85em
          \lower.35em\hbox{$\sim$}~}\xspace}
\def\lsim{{~\raise.15em\hbox{$<$}\kern-.85em
          \lower.35em\hbox{$\sim$}~}\xspace}

\def\sPlot{\mbox{\em sPlot}\xspace}

\def\pt         {\ensuremath{p_{\mathrm{T}}}\xspace}

\def\ptot       {\ensuremath{p}\xspace}

\def\tell1  {TELL1\xspace}
\def\ukl1   {UKL1\xspace}

\newcommand{\lhcborcid}[1]{\href{https://orcid.org/#1}{\hspace*{0.1em}\raisebox{-0.45ex}{\includegraphics[width=1em]{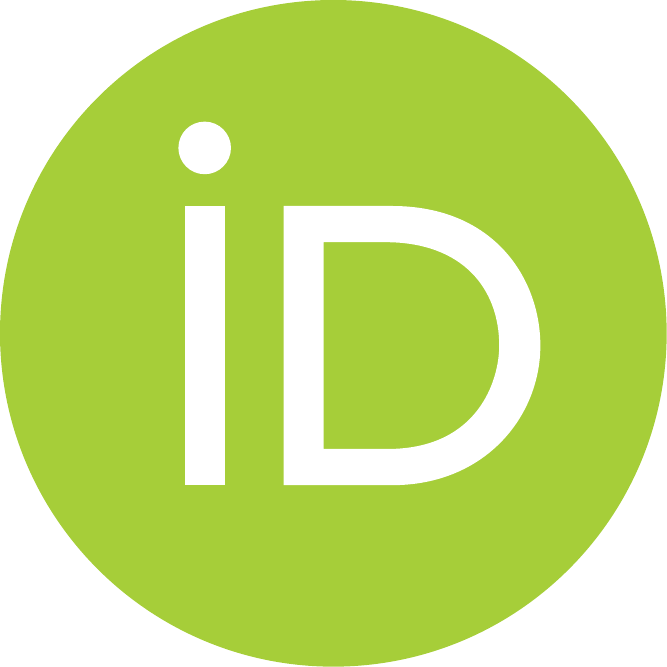}}}}

\usepackage{cite} 
\usepackage{mciteplus}

\usepackage{longtable} 

\begin{document}

\renewcommand{\thefootnote}{\fnsymbol{footnote}}
\setcounter{footnote}{1}


\begin{titlepage}
\pagenumbering{roman}

\vspace*{-1.5cm}
\centerline{\large EUROPEAN ORGANIZATION FOR NUCLEAR RESEARCH (CERN)}
\vspace*{1.5cm}
\noindent
\begin{tabular*}{\linewidth}{lc@{\extracolsep{\fill}}r@{\extracolsep{0pt}}}
\ifthenelse{\boolean{pdflatex}}
{\vspace*{-1.5cm}\mbox{\!\!\!\includegraphics[width=.14\textwidth]{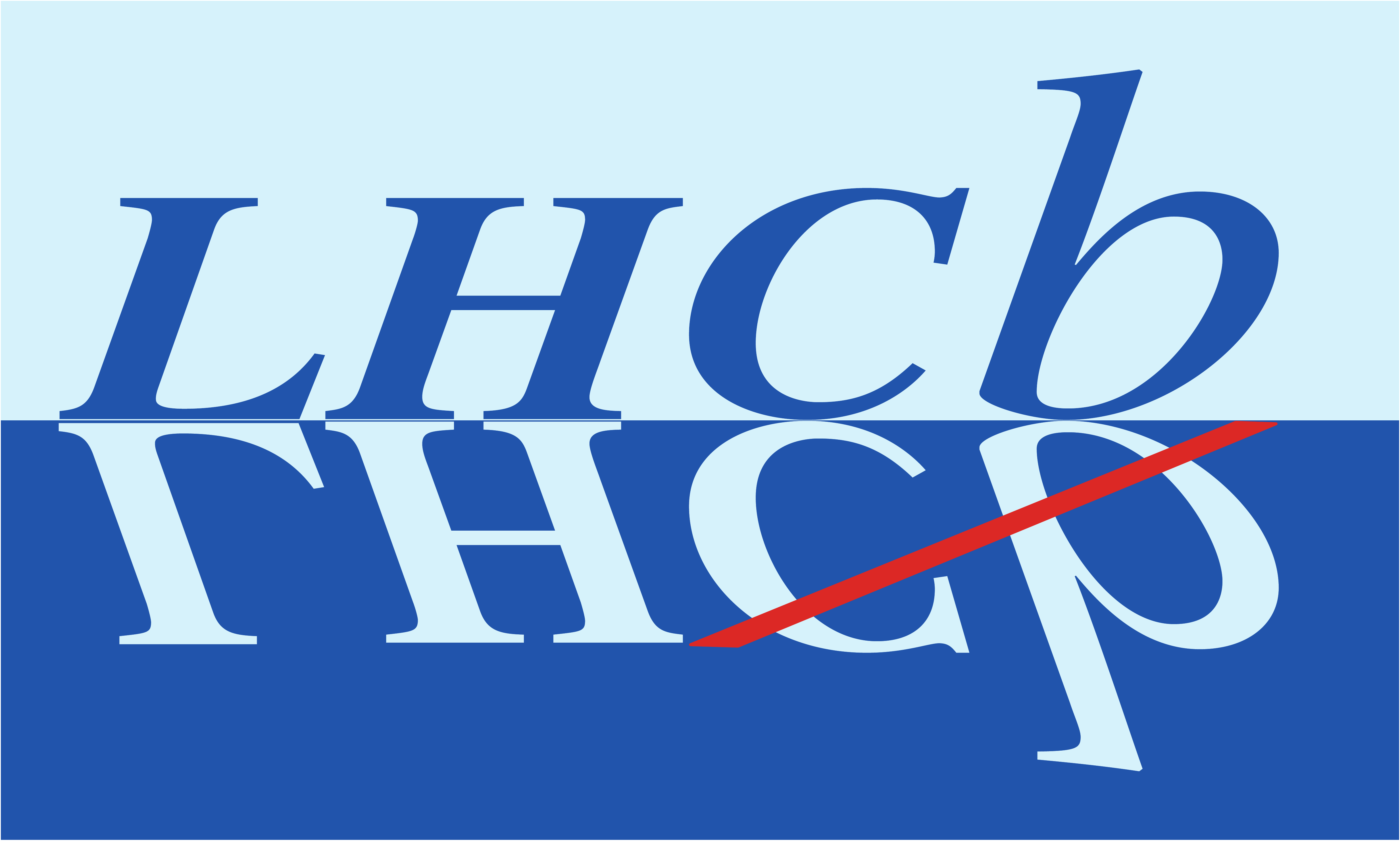}} & &}%
{\vspace*{-1.2cm}\mbox{\!\!\!\includegraphics[width=.12\textwidth]{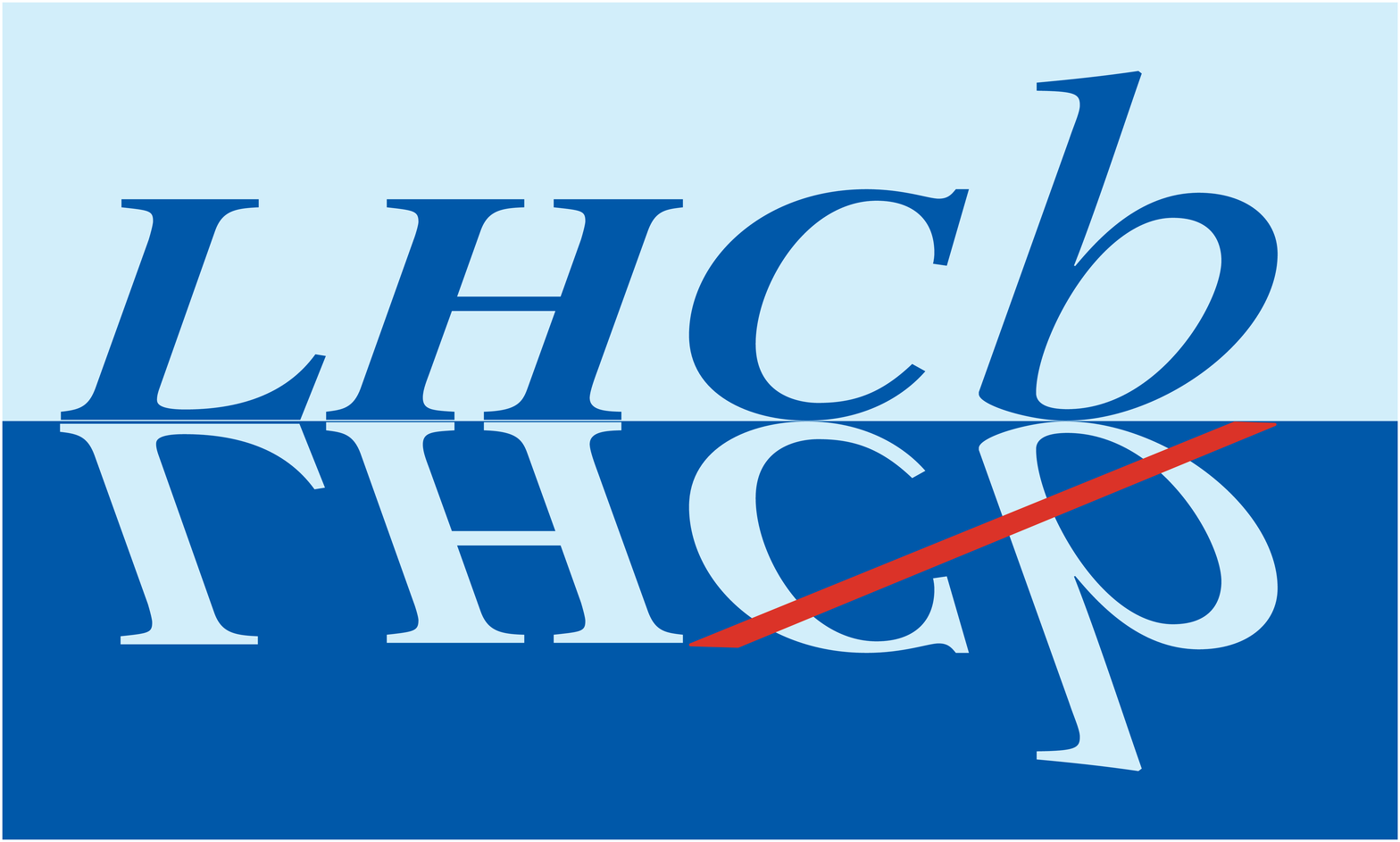}} & &}%
\\
 & & CERN-EP-2023-103 \\  
 & & LHCb-PAPER-2023-005 \\  
  & & March 20, 2024 \\
 & & \\
\end{tabular*}

\vspace*{4.0cm}

{\normalfont\bfseries\boldmath\huge
\begin{center}
  \papertitle 
\end{center}
}

\vspace*{2.0cm}

\begin{center}
\paperauthors\footnote{Authors are listed at the end of this paper.}
\end{center}

\vspace{\fill}

\begin{abstract}
  \noindent
A search for \CP violation in $\Dz \to \pim\pip\piz$ decays is reported, using $pp$ collision data collected by the \lhcb experiment from 2015 to 2018 corresponding to an integrated luminosity of 6\invfb. An unbinned model-independent approach provides sensitivity to local \CP violation within the two-dimensional phase space of the decay. The method is validated using the Cabibbo-favoured channel $\Dz \to \Km\pip\piz$ and background regions of the signal mode. The results are consistent with \CP symmetry in this decay.
  
\end{abstract}

\vspace*{2.0cm}

\begin{center}
  Published in JHEP 09 (2023) 129
\end{center}

\vspace{\fill}

{\footnotesize 
\centerline{\copyright~\papercopyright. \href{\paperlicenceurl}{\paperlicence}.}}
\vspace*{2mm}

\end{titlepage}


\newpage
\setcounter{page}{2}
\mbox{~}
%
%
%
%


\renewcommand{\thefootnote}{\arabic{footnote}}
\setcounter{footnote}{0}

\cleardoublepage


\pagestyle{plain} 
\setcounter{page}{1}
\pagenumbering{arabic}


\section{Introduction}

In the Standard Model (SM) of particle physics, \CP violation is included in the quark sector through a single weak phase in the CKM matrix~\cite{Cabibbo:1963yz,Kobayashi:1973fv}. While this satisfies one of the Sakharov conditions required for baryogenesis~\cite{Sakharov:1967dj}, the baryon asymmetry observed in the universe exceeds by several orders of magnitude the level accommodated within the SM~\cite{Huet:1994jb,Riotto:1999yt}. This motivates the search for physics beyond the SM and makes \CP-violation studies a promising direction in this search.
In the kaon and beauty sectors, \CP violation has been long established~\cite{Christenson:1964fg,KTeV:1999kad,NA48:2001bct,BaBar:2001pki,Belle:2001zzw,BaBar:2004gyj,Belle:2004nch,LHCb:2013syl,LHCb:2012xkq,LHCb:2020byh}. However, it was only in 2019 that a statistically significant observation of \CP violation was made in charm hadrons, by the \lhcb collaboration~\cite{LHCb:2019hro}. 

There is currently no clear consensus whether the observed level of \CP violation in charm hadrons can be accommodated within the SM, due to the challenges of performing calculations of asymmetries in charm decays~\cite{Buccella:2019kpn,Dery:2019ysp,Chala:2019fdb,Schacht:2021jaz,Buras:2021rdg,Bediaga:2022sxw}. 
In the SM, charm \CP violation is expected to be small due to a combination of CKM suppression and GIM cancellation. Depending on the decay mode, it is predicted to reach $\order(10^{-3})$, with the largest effects in singly Cabibbo-suppressed (SCS) channels, where \CP violation appears via interference between tree and penguin amplitudes with different strong and weak phases~\cite{Grossman,Brod:2011re,Atwood:2012ac}.
Multi-body charm decays provide a powerful tool to probe \CP violation.
In these decays, phase-space-dependent local \CP asymmetries can arise from the interference among intermediate resonances, which leads to a corresponding variation of strong phases. 
Studies of these local asymmetries provide additional sensitivity to observation of \CP violation, complementing studies of phase-space integrated asymmetries and two-body decays~\cite{Grossman:2012eb,Dery:2021mll, Qin:2013tje,Cheng:2019ggx,Cheng:2021yrn}.
 Similar investigations in the beauty sector have revealed very large local asymmetries, approaching 100\% in magnitude for some phase-space regions~\cite{LHCb:2022nyw,LHCb:2022nng}.

 In charm quark decays, \CP violation was discovered by measuring the difference in time-integrated \CP asymmetries of the two-body decays $\Dz \to \Kp\Km$ and $\Dz \to \pip\pim$, to construct the quantity $\Delta A_{\text{CP}}$~\cite{LHCb:2019hro}. Recently, a dedicated \lhcb measurement of the \CP asymmetry in $\Dz \to K^+K^-$ decays was performed. In conjunction with $\Delta A_{\text{CP}}$, this indicates that the observed \CP violation is driven by an asymmetry in the $\pi^+\pi^-$ decay mode, at greater than $3\,\sigma$ significance~\cite{LHCb:2022vcc}. 
The three-body SCS decay $\Dz \to \pim \pip\piz$ is dominated by  $\Dz \to \rho^{\pm}\pi^{\mp}$ amplitudes, which proceed via the same quark-level transitions as the decay $\Dz \to \pip\pim$.
A previous \lhcb search for local \CP violation in this decay measured a $p$-value of 2.6\% for the hypothesis of \CP symmetry~\cite{LHCb:2014nnj} and the world average of the phase-space integrated asymmetry is in agreement with \CP symmetry~\cite{HFLAV:2022pwe, BaBarAcp, Belle:2008qrk, CLEO:2005uoz}.
The third most important amplitude of the $\Dz \to \pim \pip\piz$ decay, $\Dz \to \rho^{0}\pi^{0}$, shares the weak structure with the two-body decay $\Dz \to \pi^{0}\pi^{0}$, whose world-average \CP asymmetry is also in agreement with \CP symmetry~\cite{HFLAV18, Belle:2014evd, CLEO:2000opx}.

In this paper, a search for local \CP violation in the decay ${\Dz \to \pim\pip\piz}$ is performed using \lhcb data collected during LHC Run~2 from 2015--2018. An unbinned model-independent method called the energy test~\cite{zech1,zech2,Williams:2011cd,Parkes:2016yie} is used to determine a $p$-value quantifying compatibility with the null hypothesis of \CP symmetry. The method follows that used in the predecessor \lhcb analysis~\cite{LHCb:2014nnj}.

\section{Analysis overview}

The energy test~\cite{zech1,zech2,Williams:2011cd,Parkes:2016yie} is an unbinned statistical test to compare two distributions and quantify their agreement with being drawn from a common source, making it appropriate for comparing signal distributions in multidimensional phase space of particle decays. The test has been used in several previous LHCb searches for \CP violation~\cite{LHCb:2014nnj, LHCb:2016qbq, LHCb:2019oke}. In this analysis the energy test is used to compare samples of $\Dz$ and $\Dzb$ mesons decaying to the $\pi^-\pi^+\pi^0$ final state. 
 The test is insensitive, by construction, to any global charge asymmetries present in the sample. It provides a single $p$-value quantifying consistency with the \CP symmetry hypothesis. 
 This analysis benefits from several improvements in the technique, which reduce the computational requirements and hence enable the test to be applied to much larger data samples than was previously feasible~\cite{Barter:2018xbc,Zech:2018bpy,Gillam:2018ufy}.

A sample of flavour-tagged \decay{\Dz}{\pim\pip\piz} decays is obtained by selecting \Dstarp candidates that decay via \decay{\Dstarp}{\Dz(\ra\pim\pip\piz)\pip}.\footnote{Unless stated otherwise, the charge conjugate mode is implicit throughout.} The charge of the low-momentum (soft) pion from the $D^{*+}$ decay is used to identify the flavour of the accompanying \Dz candidate. The analysis methods are developed and validated using three independent approaches: realistic pseudoexperiment simulations with and without \CP asymmetries injected; studies of a topologically similar control mode in collision data, where \CP violation is expected to be negligible; and studies with the signal mode following randomisation of the charm meson flavour and injection of fake asymmetries to mimic nuisance effects from production and detection asymmetries. After all methods are validated the energy test is performed in the signal channel to search for \CP asymmetries. 

For a three-body meson decay with spinless initial and final states, the phase space is fully defined by two kinematic quantities. In this analysis, the Dalitz-plot formalism is used to describe the signal decay using squared invariant masses of two-body combinations of final state particles. To enable a direct comparison of the Dalitz plane distributions for $\Dz$ and $\Dzb$ decays, a set of flavour-symmetric coordinates is defined: $s_{12} \equiv m^2(\pi^{\pm}\pi^{\mp})$, $s_{13} \equiv m^2(\pi^{\pm}\pi^0)$, and $s_{23} \equiv m^2(\pi^{\mp}\pi^0)$, where in all cases the upper (lower) charge superscript refers to the $\Dz$ ($\Dzb$) case. 

To assess the impact of experimental effects such as production and detection asymmetries, the Cabibbo-favoured (CF) decay $\Dz \rightarrow K^- \pi^+ \pi^0$ is used as a control channel, which is topologically and kinematically similar to the signal decay. The CF channel benefits from a yield approximately eight times greater than the signal. This enables the control mode to be split into several independent samples with signal-like yields to test for biases from detector-induced asymmetries. Wherever possible, the event selection for the control mode is chosen to match the corresponding signal selection.
The main difference between the signal and control modes is the presence of a charged kaon in the control mode. Due to material effects in the detector, $K^-\pi^+$ detection asymmetries are known to be much larger than for $\pi^+\pi^-$~\cite{LHCb-PAPER-2014-013}, so the signal mode is anticipated to suffer from smaller nuisance asymmetries than the control mode. 

\section{LHCb experiment}

The \lhcb detector~\cite{LHCb-DP-2008-001,LHCb-DP-2014-002} is a single-arm forward spectrometer covering the \mbox{pseudorapidity} range $2<\eta <5$, designed for the study of particles containing \bquark or \cquark
quarks. The detector includes a high-precision tracking system consisting of a silicon-strip vertex detector surrounding the $pp$ interaction region~\cite{LHCb-DP-2014-001}, a large-area silicon-strip detector located upstream of a dipole magnet with a bending power of about $4{\mathrm{\,Tm}}$, and three stations of silicon-strip detectors and straw drift tubes~\cite{LHCb-DP-2017-001} placed downstream of the magnet. The tracking system provides a measurement of the momentum, \ptot, of charged particles with a relative uncertainty that varies from 0.5\% at low momentum to 1.0\% at 200\gevc. The minimum distance of a track to a primary $pp$ collision vertex (PV), the impact parameter (IP), is measured with a resolution of $(15+29/\pt)\mum$, where \pt is the component of the momentum transverse to the beam, in\,\gevc. Different types of charged hadrons are distinguished using information from two ring-imaging Cherenkov detectors~\cite{LHCb-DP-2012-003}. Photons, electrons and hadrons are identified by a calorimeter system consisting of scintillating-pad and preshower detectors, an electromagnetic and a hadronic calorimeter. Muons are identified by a
system composed of alternating layers of iron and multiwire proportional chambers~\cite{LHCb-DP-2012-002}. The online event selection is performed by a trigger~\cite{LHCb-DP-2012-004}, which consists of a hardware stage based on information from the calorimeter and muon systems, followed by a software stage which applies a full event reconstruction.

The neutral pion candidate in the \decay{\Dz}{\pim\pip\piz} final state is reconstructed from its diphoton decay based on energy deposits in the electromagnetic calorimeter.
The reconstructed neutral pions can be separated into two categories, depending on the opening angle between the photons produced in the \decay{\piz}{\gamma\gamma} decay. A \piz candidate is labelled \textit{resolved} if the decay photons are reconstructed as two distinct electromagnetic calorimeter clusters and as \textit{merged} if the two photons are reconstructed as a single cluster~\cite{LHCb-DP-2014-002}. The two categories differ mainly in the kinematics of the neutral pion (merged $\pi^0$ candidates tend to have higher momentum than resolved) and therefore the reconstructed Dalitz distributions for the two types are also quite different. To maximise the sensitivity to potential \CP violation effects over the entire phase space, the two categories are combined before applying the energy test. However, the event selection is optimised separately for the two categories to reflect their different characteristics.

\section{The energy test method}
\label{sec:etmethod}

In this analysis, the energy test method is used to compare the Dalitz distributions of \Dz and \Dzb meson decays to provide a sensitive search for \CP violation within the phase space of these decays. The test quantifies localised sample differences by calculating a single test statistic defined as
\begin{equation*}
    \tsymbol \equiv  \dfrac{1}{2n(n-1)}\sum_{i,j\neq i}^{n}\psi_{ij}
+ \dfrac{1}{2\overline{n}(\overline{n}-1)}\sum_{i,j\neq i}^{\overline{n}}\psi_{ij}
- \dfrac{1}{n\overline{n}} \sum_{i,j}^{n,\overline{n}}\psi_{ij}.
\end{equation*}
The function $\psi_{ij}$ depends on the distance between pairs of candidates in the phase space, $d_{ij}$, such that $\psi_{ij} =  \psi(d_{ij})$. The first double summation is over all possible pairs of ($n$) $\Dz$ candidates, the second is over all pairs of ($\overline{n}$) $\Dzb$ candidates, while the third is over all $\Dz$-$\Dzb$ pairs.
The normalisation of each term removes any effects from global asymmetries, such that by construction only local asymmetries are probed in the energy test. Under the null hypothesis that the two samples are drawn from the same underlying distribution, the \tvalue is drawn from a distribution around zero.  In the presence of differences between the two samples, as might be caused by \CP violation~\cite{zech1,zech2,Williams:2011cd,Parkes:2016yie}, the \tsymbol distribution shifts to larger values. 

For this analysis, $\psi_{ij}$ is chosen to be a Gaussian function $\psi_{ij}=e^{-d_{ij}^2/2\delta^2}$, and the distance between candidate pairs is defined symmetrically in terms of the three squared invariant masses of two-pion combinations, $d_{ij}^2 = [(\Delta s_{12})^2_{ij} + (\Delta s_{13})^2_{ij} + (\Delta s_{23})^2_{ij}]$, where $(\Delta s)_{ij}$ gives the difference in the coordinate $s$ for candidates $i$ and $j$.
The tunable parameter $\delta$ constrains the effective radius in phase space within which local asymmetries can be probed. Its value is optimised to maximise sensitivity to \CP violation as discussed in Sec.~\ref{sec:sensitivity}. 

The calculated \tvalue can be converted into a $p$-value by comparing it to the distribution expected under the null hypothesis. This distribution is determined by running the test repeatedly where candidates of both flavours are randomly assigned to the two samples. The fraction of these permutation tests with $T$-values greater than the value under test then sets the $p$-value of the overall test. 
The distribution of $T$ under the null hypothesis scales directly with the sample size~\cite{Barter:2018xbc,Zech:2018bpy,Gillam:2018ufy}, so can be determined using significantly smaller samples than used for the nominal result.

\section{Event selection}

This analysis uses proton-proton collision data corresponding to an integrated luminosity of approximately 6\invfb, collected during the LHC Run~2 period (2015--2018) at a centre-of-mass energy $\sqrt{s} = 13$~TeV.

At the hardware trigger stage, signal \decay{\Dz}{\pim\pip\piz} events are selected either through significant energy deposits in the hadronic calorimeter (associated with one or both of the charged pions from the $\Dz$ decay), or through deposits in the electromagnetic calorimeter consistent with a photon shower (associated with the neutral pion decay). Other events passing the hardware trigger are also selected if the trigger decision is independent of the signal candidate.

 At the first stage of the software trigger a partial event reconstruction is performed. Candidates are selected if they satisfy either of two multivariate displaced track algorithms. The first accepts a single track if it fulfils a two-dimensional requirement for minimum values in (\pt, $\chi^2_{\text{IP}}$) space, where $\chi^2_{\text{IP}}$ is the difference in the $\chi^2$ of the PV fit with and without the track under consideration. This algorithm preferentially accepts tracks with large values of $\pt$, $\chi^2_{\text{IP}}$, or both. The second is a two-track algorithm utilising several variables within a boosted decision tree, sensitive to track pairs produced at a significant distance from the PV.

At the second software trigger stage, a full event reconstruction is performed taking into account the best available knowledge of detector alignment and calibration. Candidates must pass either dedicated channel-specific triggers, or an inclusive topological trigger. The latter exploits the characteristic signature $D^{*+} \to \Dz\pi^+$ with $\Dz \to h^+h^{'-}X$, where $h$ represents any charged long-lived hadron and $X$ indicates unreconstructed decay products. 

Following the trigger stage, the offline selection starts with the reconstruction of \Dz candidates. The two charged pions from the $\Dz$ decay must be significantly displaced from the PV and fulfil particle identification requirements to remove backgrounds from kaons and incorrectly reconstructed tracks (ghosts). The two pions must form a good-quality vertex and have an invariant mass below 1850~MeV/$c^2$. The neutral pion must be identified using dedicated LHCb algorithms via resolved or merged photon signatures, and the diphoton system must have an invariant mass consistent with the known $\pi^0$ mass. A $\Dz$ candidate is then reconstructed from the combination of the neutral pion and two charged pions. The three-particle system must have an invariant mass consistent with the known mass of the $\Dz$ meson.
The direction of the $\Dz$ momentum vector is required to be consistent with that of the displacement vector joining the production and decay vertices.

To reconstruct the $D^{*+}$ candidate, an additional charged pion is combined with the $\Dz$ meson. This soft pion is subject to particle identification requirements to remove backgrounds from ghost tracks and misidentified particles, and must be consistent with originating at the PV. The $\Dz\pi^+$ system is required to have a good vertex quality and a reconstructed mass difference $\Delta m \equiv m(D^{*+})-m(\Dz)$ less than 180~MeV/$c^2$.

When reconstructing the decay a kinematic refit of the full $D^{*+}$ decay chain is performed to improve the resolution with externally imposed constraints~\cite{Hulsbergen:2005pu}. The $D^{*+}$ candidate is constrained to originate from the PV, and the $\pi^{0}$ candidate is mass-constrained to the known experimental value. When calculating the Dalitz variables used as inputs to the energy test, there is an additional requirement constraining the reconstructed mass $m(\pi^{-}\pi^{+}\pi^{0})$ to the known $D^{0}$ mass, which ensures that candidates lie in the physically allowed kinematic region. 

Potential contamination from clone tracks, corresponding to single charged particles reconstructed as multiple near-colinear tracks, is efficiently suppressed by requiring that all pairs of tracks have an angular separation exceeding 0.0005 rad. 

After the offline selection procedure, the dominant background remaining in the sample corresponds to random combinations of tracks and neutral pions. This background is significantly higher for the resolved $\pi^0$ channel, due to the low-$\pt$ diphoton signature. To further suppress this background a multivariate classifier is developed and applied to the resolved channel alone. The figure-of-merit to be maximised is the quantity $S/\sqrt{(S+B)}$, where $S$ ($B$) is the signal (background) yield within the signal region defined by $|\Delta m - 145.4| < 1.8{\mevcc}$. 
The multivariate selection gives an appreciable gain in signal significance for the resolved channel, while no corresponding improvement is found when applying a similar approach for the merged channel.

A boosted decision tree (BDT) classifier is trained and tested using data, based on the TMVA framework~\cite{tmva}. The signal (background) sample is formed of \mbox{$\Dz \to \pim \pip \piz$} candidates each with a signal  (background) weight determined using the \sPlot technique~\cite{sPlotsPaper} with  $\Delta m$ as the discriminating variable.
To avoid biases from the BDT training, the data are divided into two disjoint halves such that the classifier trained on one half is applied to the second half and vice versa.

After applying all selection criteria, a small fraction of events contain more than one signal candidate. These may affect the present analysis, as they can lead to candidates with identical or near-identical phase-space coordinates. As a precaution to mitigate potential biases in the energy test, a single candidate is randomly chosen to be retained in such cases. This affects approximately 3\% (6\%) of events in the merged (resolved) samples.

\begin{figure}[!t]
\centering	 	        
\includegraphics[width=0.5\textwidth]{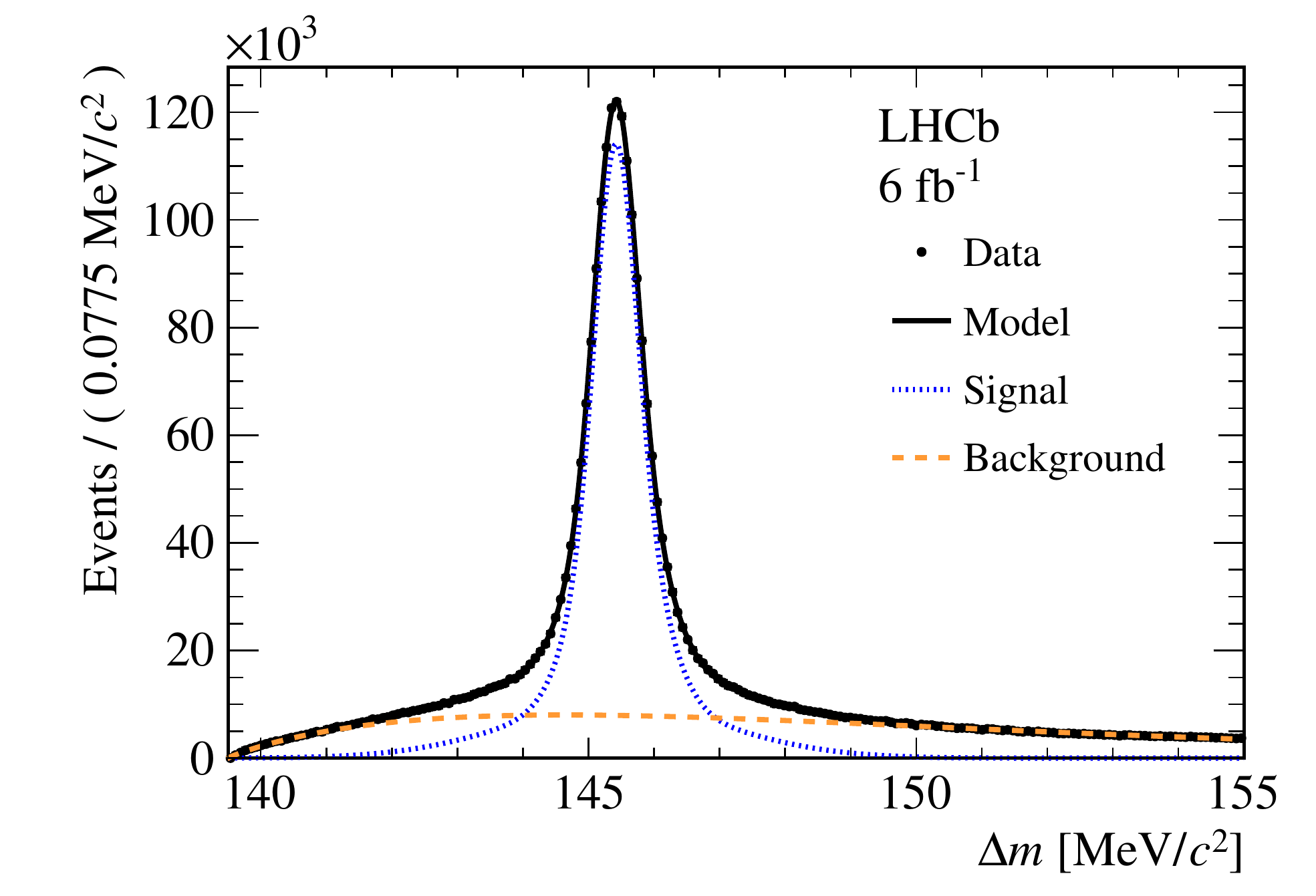}\hfill
\includegraphics[width=0.5\textwidth]{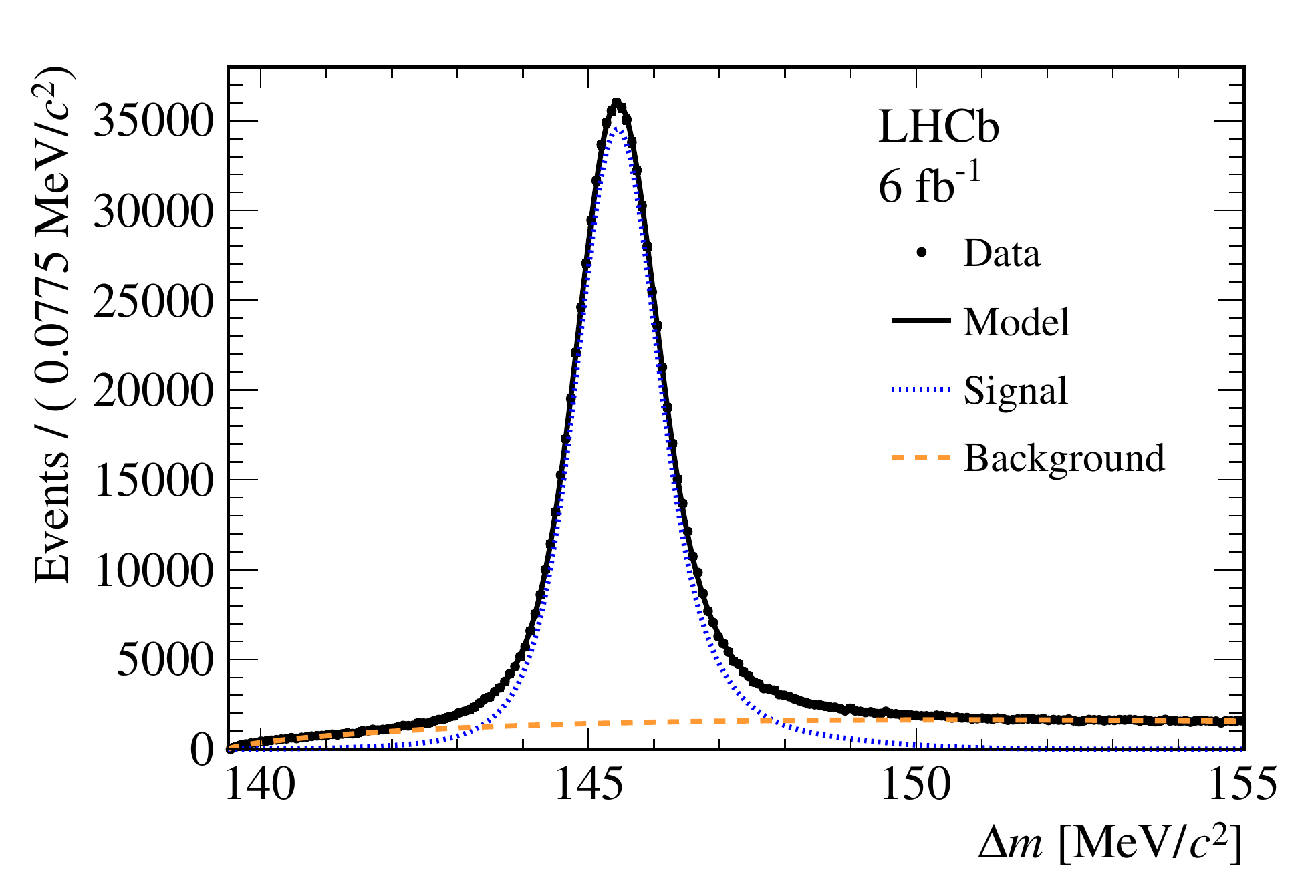}\hfill
\caption{ Distributions of $\Delta m$ for (left) the resolved and (right) merged $\pi^0$ samples after the application of all selection criteria except the $\Delta m$ requirement. Also shown are projections of the fit model used to evaluate signal yields and purity.}
\label{fig:Final_Fits}
\end{figure}

The $\Delta m$ distributions after applying all selection criteria are shown in Fig.~\ref{fig:Final_Fits}. These distributions are fitted to obtain signal and background yields for use in sensitivity studies, as described later.
For these fits the signal peaks are modelled by the sum of three Gaussian functions, where for the resolved sample one of them has different width parameters for the left and right halves of the peak. The background is described by an empirical threshold shape $F_{{\rm bkg}}(x) = (x - x_0)^\alpha e^{-\beta(x - x_0)}$, with $\alpha$ and $\beta$ freely varying in the fit and with $x_0$ fixed to the known mass of the charged pion. All free parameters for both signal and background models are independent for the merged and resolved samples.

When applying the energy test, only candidates within the region ${|\Delta m - 145.4| < 1.8{\mevcc}}$ are used. This yields a sample of 1.71M (0.76M) signal decays with a purity of 81\% (91\%) for the resolved (merged) sample. This exceeds by more than a factor of four the signal yield of the corresponding LHCb Run~1 analysis~\cite{LHCb:2014nnj}. 

The control channel $D^{0} \rightarrow K^{-}\pi^{+}\pi^{0}$ is selected using the same criteria except for the particle identification requirements on the kaon and pion, which result in mutually exclusive samples for the signal and control modes.
The kinematic distributions of the final state particles are similar for the selected signal and control mode data. 
The yield of the control mode is 5.32M (13.78M) for the merged (resolved) $\pi^0$ samples, with a purity of 97\% (94\%). 

To suppress the potential impact of nuisance asymmetries, 
the number of candidates collected in each of the two LHCb dipole magnet polarities is equalised by randomly rejecting candidates from the larger sample. This mitigates against potential instrumental asymmetries due to geometrical effects, whereby particles of a particular charge are more likely to be swept out of the acceptance. The equalisation improves the cancellation of such effects at the cost of around 5\% reduction in signal yield. 
%
%
The equalisation is enforced independently for each of the four data-taking years (2015--2018) to reflect different conditions, for both signal and control modes. 

\begin{figure}[!t]
\centering	 	        
\includegraphics[width=0.49\textwidth]{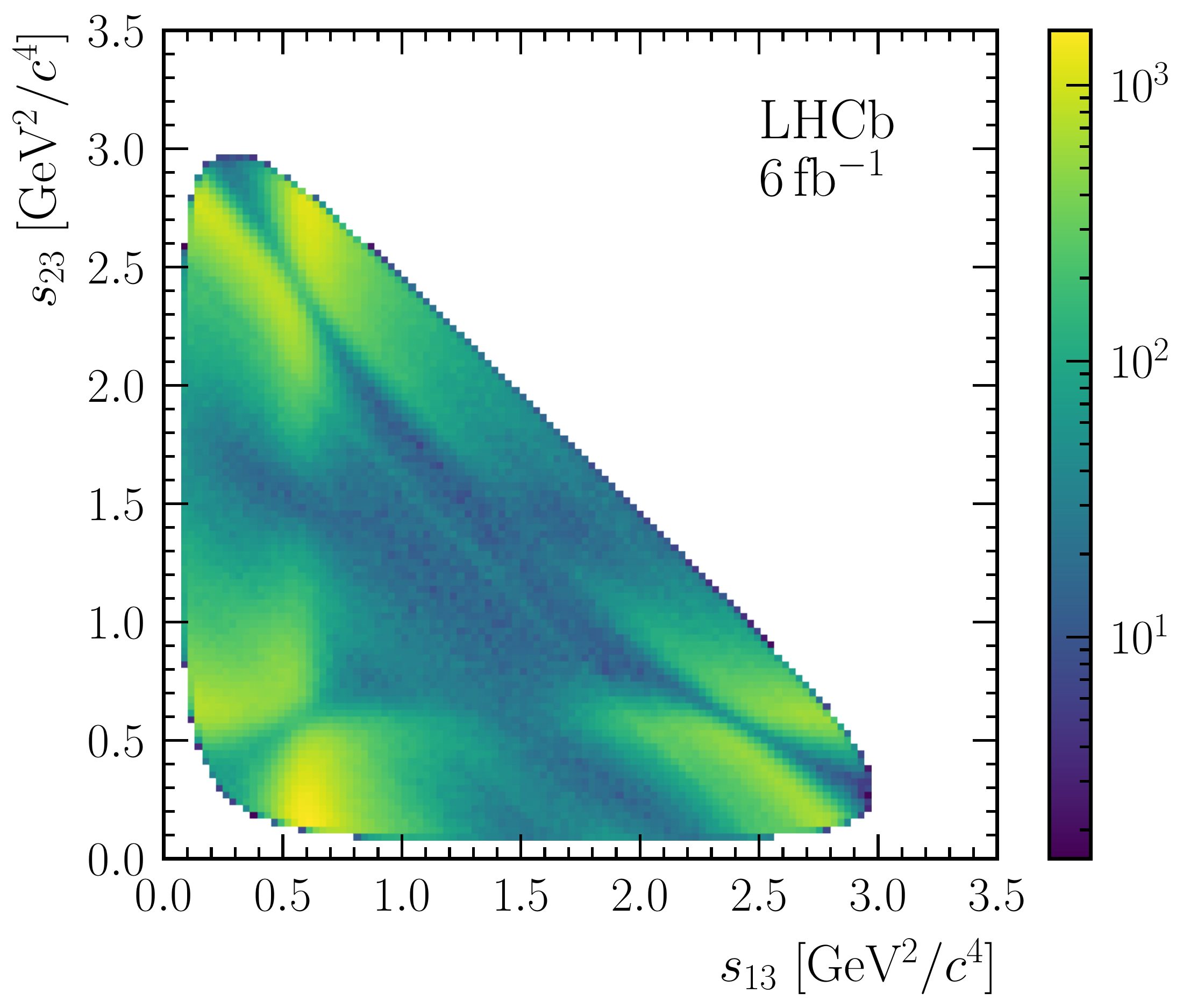}
\hfill
\includegraphics[width=0.49\textwidth]{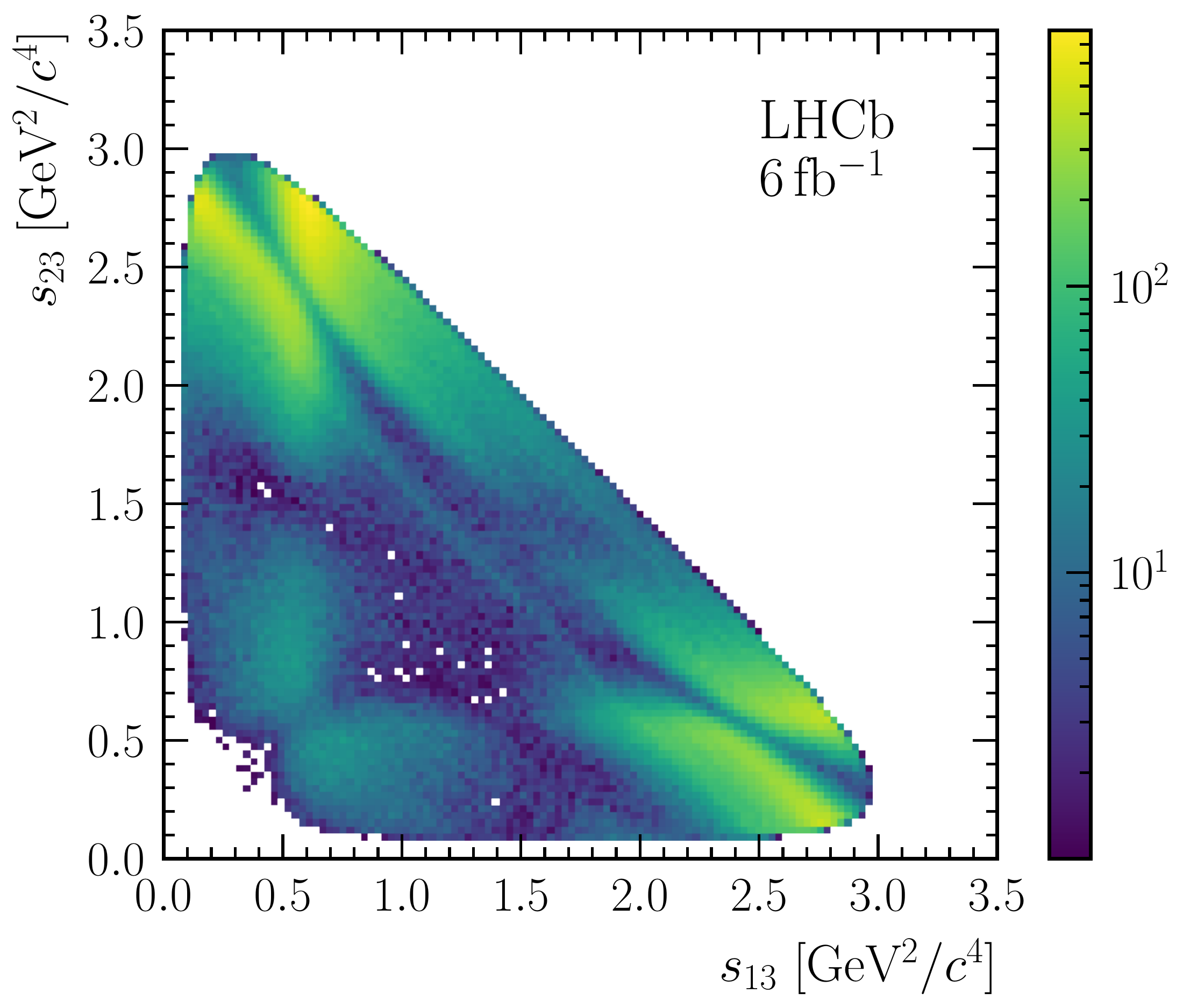}
\hfill
\caption{Dalitz plots for the background-subtracted signal samples for (left) the resolved and (right) merged $\pi^0$ categories, with the two $m^2(\pi^{\pm}\pi^0)$ variables chosen for the projection. The three $\rho^{\pm,0}$ resonances dominate the phase space.}
\label{fig:DalitzPlots_sWeight}
\end{figure}

The background-subtracted Dalitz distributions for the resolved and merged $\pi^0$ categories are shown in Fig.~\ref{fig:DalitzPlots_sWeight}. Three $\rho^{\pm,0}$ resonances are clearly visible as two-lobe structures in the Dalitz plane. These dominate the phase space, as expected from previous studies~\cite{3pi_babar_model}. Candidates with large values of $s_{12}$, corresponding to the lower left corner of the Dalitz plane, have little remaining centre-of-mass energy for the neutral pion. As such, this region of phase space is dominated by resolved (low momentum) $\pi^0$ candidates. 

\section{Sensitivity studies and optimisation}
\label{sec:sensitivity}

The energy test has a single tunable parameter, $\delta$, which sets the phase-space scale of the test and is chosen to optimise the sensitivity of the method. The optimal value depends on the size of the dataset and the distribution of candidates in the phase space. Different values chosen for $\delta$ may be optimal for different \CP violation scenarios. Pseudoexperiments are used to study such effects and select the $\delta$ value for this analysis.

In the pseudoexperiments, phase-space distributions of signal decays are generated using the \textsc{Laura}$^{++}$ package~\cite{laura++}, with an amplitude model taken from a previous analysis of this channel~\cite{3pi_babar_model}. This model consists of several components from different intermediate resonances and a non-resonant term, each multiplied by a complex coefficient. The dominant contributions are from charged and neutral $\rho(770)$ resonances, with fit fractions of 67.8\% ($\rho^+\pi^-$), 34.6\% ($\rho^-\pi^+$), and 26.2\% ($\rho^0\pi^0$) for the \Dz decay. 
The $\Dz$ and $\Dzb$ complex coefficients, either in their magnitudes or phases, are adjusted for both the dominant ($\rho(770)^+\pi^-$) and sub-dominant ($\rho(770)^-\pi^+$) intermediate states to emulate \CP-violation.

For each pseudoexperiment the number of signal candidates is set to match the corresponding signal yield in data, for the combination of merged and resolved samples. Nonuniform efficiency effects and background contamination are implemented based on studies with the data to make the pseudoexperiments more realistic. The efficiency variation across the Dalitz plane is estimated by comparing background-subtracted data with the generator-level simulation. It is found to vary smoothly over the Dalitz plane, with around 10\% variation across the two $m^2(\pi^{\pm}\pi^0)$ variables, and decreasing by around 50\% from low to high $m^2(\pi^+\pi^-)$. For different studies, backgrounds are generated either with or without a global charge asymmetry.

\begin{figure}[!t]
    \centering
    \includegraphics[width=0.7\textwidth]{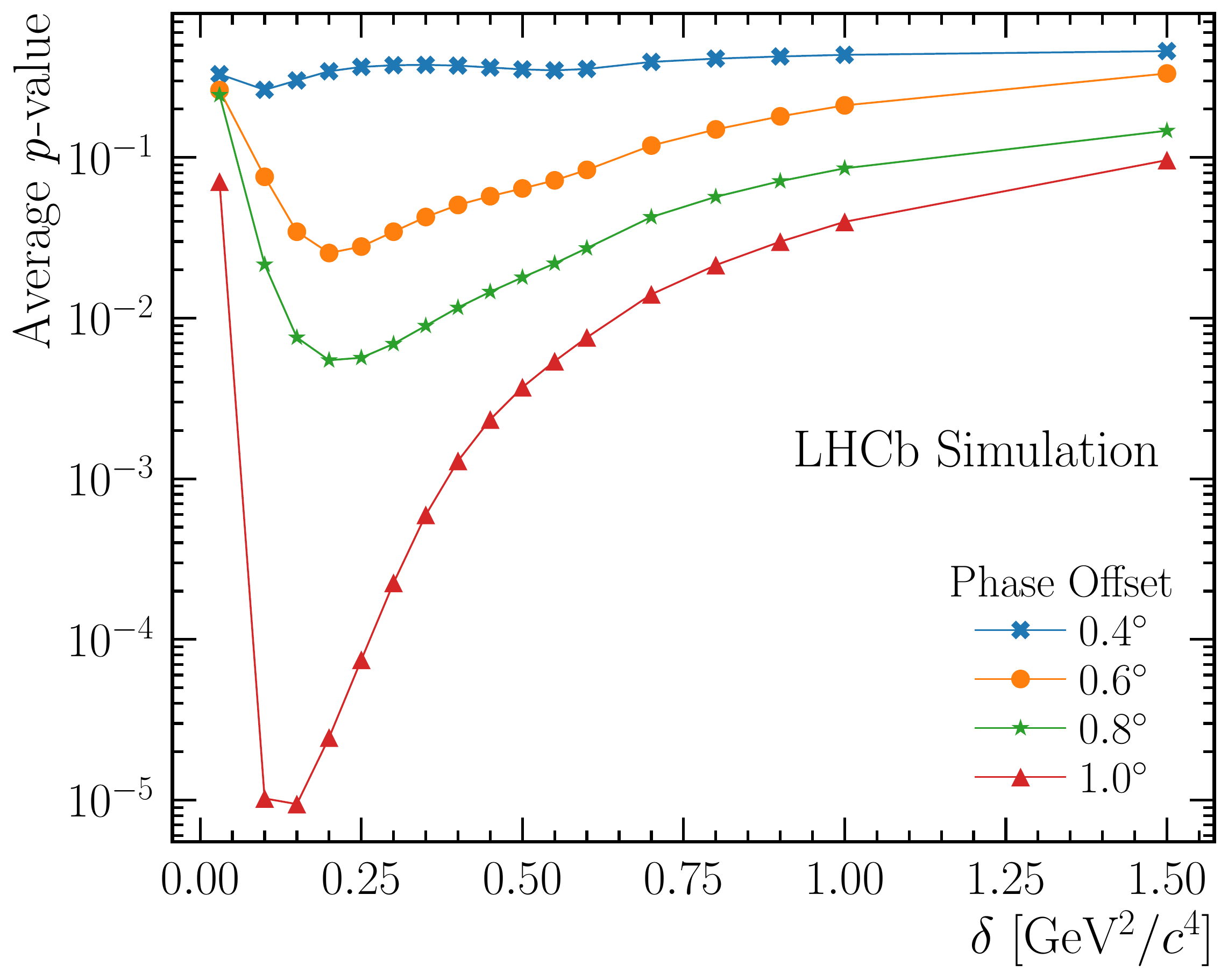}
    \caption{Average $p$-value as a function of the energy test distance parameter $\delta$, for different values of the \Dz-\Dzb phase difference for the \decay{\Dz}{\rho(770)^{-}\pi^+} amplitude. Straight guide lines are also drawn connecting successive points for each scenario.}
    \label{fig:realistic_subdom_pvals}
\end{figure}

Pseudoexperiments are generated with \CP asymmetries of varying size to study the sensitivity of the analysis as a function of the choice of $\delta$ value. Phase (magnitude) differences between $\Dz$ and $\Dzb$ decay amplitudes are generated in the range $0.1\degree$--$1.0\degree$ (0.1\%--1.0\%). These pseudoexperiments are evaluated using the energy test method at 17 different values of $\delta$ ranging from 0.03--1.5~$(\text{GeV}/c^2)^2$. In each case a $p$-value is determined and these are plotted as a function of $\delta$. The optimal choice of $\delta$ for a given pseudoexperiment is that which gives the smallest $p$-value. To reduce the impact of statistical fluctuations, five pseudoexperiments are generated for each \CP violation scenario and the average $p$-value is calculated for each point in the scan. An example of this procedure is presented in Fig.~\ref{fig:realistic_subdom_pvals}, which shows the $\delta$ scans for different phase differences injected into the sub-dominant \decay{\Dz}{\rho(770)^-\pi^+} amplitude. The minimum $p$-value is found close to $\delta=0.2\,(\text{GeV}/c^2)^2$ for all cases, hence this is chosen as the distance parameter for the energy test throughout this analysis.

To assess the overall sensitivity for a particular \CP asymmetry scenario, an ensemble of 500 pseudoexperiments is generated with a phase difference of 0.9\degree\ in the dominant \decay{\Dz}{\rho(770)^+\pi^-} amplitude. The significance of the result, expressed as a one-sided Gaussian $z$-score, follows a Gaussian distribution with a mean of $4.3\,\sigma$ and a width of $0.7\,\sigma$. If nature exhibited a \CP asymmetry of this type, this analysis would obtain a $3\,\sigma$ evidence in 95\% of cases.
Similar sensitivity studies indicate that the energy test applied to the current sample is insensitive to \CP phase asymmetries below 0.5\degree and to magnitude asymmetries below 0.5\%, for the two specific amplitudes investigated.
Should a significant \CP asymmetry be observed, a more complete amplitude analysis would be needed to fully understand the results.

\section{Validation and cross-checks using data}

Several checks are performed with data to assess the potential impact of nuisance asymmetries on this analysis and confirm that the reported $p$-value is unbiased. This section describes studies using the control mode, background-dominated samples, and pseudoexperiments produced with the signal data sample with flavour information removed and different nuisance asymmetries injected.
These studies demonstrate that any local nuisance asymmetries which might affect the signal decay mode are sufficiently small that they can be neglected without impacting the results. There are global asymmetries of order 1\%, but these do not vary significantly over the Dalitz plane. 

For the control mode $\Dz \to K^-\pi^+\pi^0$, the combined sample with both $\pi^0$ meson types contains approximately 19 million candidates after all selection requirements are imposed. The fraction of merged $\pi^0$ candidates is 27\%, consistent with the 26\% fraction for the signal mode. 
The energy test is evaluated on eight independent, randomly chosen subsamples of the control mode data, each having a yield approximately matching that in the signal channel. 
The eight resulting $p$-values are consistent with having been drawn from a uniform distribution between zero and one, indicating that any nuisance asymmetries from production or detector effects are small enough to be neglected in the signal channel.

Here and elsewhere, the agreement of the set of $p$-values, $p_i$, with being drawn from a uniform distribution is tested by calculating the statistic $P = -2 \sum_{i=1}^{N} \log p_i$. The $P$-statistic follows a $\chi^2$ distribution with $2N$ degrees of freedom if the individual $p$-values are sampled from a uniform distribution. Hence, a value of $P$/2$N$ significantly greater than unity would be indicative of a nonuniform distribution. 

For the eight-sample test over the control mode, no evidence is found of any deviation from a uniform $p$-value distribution, with $P/2N = 16.4/16$. A further test is performed with a division into four subsamples, each with twice the signal yield. Again, no indication of non-uniformity is observed ($P/2N = 10.3/8$). This confirms that this method is insensitive to existing nuisance asymmetries even with double the sample size currently available.

Charge asymmetries in background candidates may be present and can in principle bias the energy test if they vary across the phase space. To supplement the control mode checks in the signal $\Delta m$ region, the energy test is also performed in background-dominated $\Delta m$ sidebands for both signal and control modes. This region is defined as the union of the ranges $(\Delta m < 142{\mevcc})$ and $(150 < \Delta m < 155{\mevcc})$. Evaluating the energy test using the 810k candidates in this region of the signal channel gives a $p$-value of 0.57. In the control mode a set of three sideband samples are defined, each with a similar yield to the signal mode sideband sample, with none giving a $p$-value below 0.2 from the energy test. Hence, there is no indication that background asymmetries are biasing this analysis, with the current sample sizes.  

To supplement the above studies, a suite of checks are performed using the signal sample. Flavour-randomised pseudoexperiments are produced by randomly labelling each signal candidate as $\Dz$ or $\Dzb$. These \CP-symmetric pseudoexperiments are then injected with nuisance asymmetries of various types, by suitably rejecting candidates according to their assigned flavour. By running the energy test over large ensembles of such samples, the potential bias from these nuisance asymmetries on the measured $p$-value can be quantified with higher confidence than the control mode studies, which are limited to at most eight independent subsamples.

For each type of nuisance asymmetry to be investigated, the first step is to use external inputs to construct binned asymmetry maps as a function of the signal particle kinematics. These asymmetries are then injected into the flavour-randomised samples.  
The realistic nuisance asymmetries are determined using a sample of $\Dstarp \to \Dz (\to \KS\pi^-\pi^+)\pi^+$ decays, where no significant \CP violation is expected. The selection of this channel is aligned where possible with the $\Dz \to \pi^-\pi^+\pi^0$ signal mode, resulting in around $12$M signal candidates with 96\% purity within the selected $\Delta m$ range.
From this sample, the raw asymmetry in $\Dz$ versus $\Dzb$ candidates is determined as a function of the reconstructed kinematic variables for different particles.

Nuisance asymmetries are determined as a two-dimensional map of the soft pion transverse momentum and pseudorapidity. While a significant global asymmetry is observed of around $-0.9\%$, there is little indication of any trend across the kinematic distributions. Using this map to generate an ensemble of asymmetry-injected flavour-randomised signal samples, and running the energy test over each one, a $p$-value distribution is obtained and found to be consistent with being uniform between zero and one (${P/2N = 98/100}$). This study is extended to inject polarity-specific asymmetries and again the $p$-values are consistent with being uniformly distributed ($P/2N = 205/200$). 

Injecting asymmetries as a function of the soft pion kinematics, as described above, is found to account for possible asymmetry dependence on kinematics of other particles in the decay chain. This is checked by comparing appropriate asymmetry maps from the $\Dz \to \KS\pi^-\pi^+$ channel with those from pseudoexperiments injected with asymmetries based on soft pion kinematics alone. As an additional check, asymmetries are injected as a function of $(\pt,\eta)$ for the two pions from the $\Dz$ decay. Again, no evidence is found for any bias in the $p$-value evaluation. 

In addition to checking $p$-value distributions of independent pseudoexperiments for uniformity, the same procedure is used to assess potential biases in the $p$-value from a single pseudoexperiment, when it is repeatedly and independently subjected to the asymmetry injection. For such checks, the significance ($z$-score) for the flavour-randomised sample is compared to the distribution of significances obtained after injecting asymmetries. This process is repeated several times to assess potential biases. The observed shifts in significance are generally consistent with Gaussian distributions centred at zero, with a few cases where a small but significant mean shift is observed of up to $0.1\,\sigma$ (in both directions). The overall conclusion of these studies is that the energy test is robust against any potential nuisance asymmetries in the signal sample.

As further cross-checks, the energy test is evaluated separately using subsamples of the control mode defined according to the data-taking year and in three disjoint categories according to the type of hardware trigger satisfied by the event. Subsamples are also defined by splitting into ranges of soft pion or $\Dz$ transverse momentum or pseudorapidity. No significant $p$-values are obtained, except when requiring that the hadronic hardware trigger is satisfied by one or both of the charged hadrons ($K^-$ or $\pi^+$) from the $\Dz$ decay. In that case $p$-values below 1\% are found for both such subsamples with signal-like yields. If the category is redefined such that the charged pion must by itself satisfy the hadronic trigger, no significant $p$-values are obtained. Hence we conclude that this bias is caused by the charged kaon, which is known to be associated with larger nuisance asymmetries from material interaction effects, and which is absent in the signal mode. 

\section{Results and conclusions}

\begin{figure}[!tb]
\centering
\includegraphics[width=0.7\textwidth]{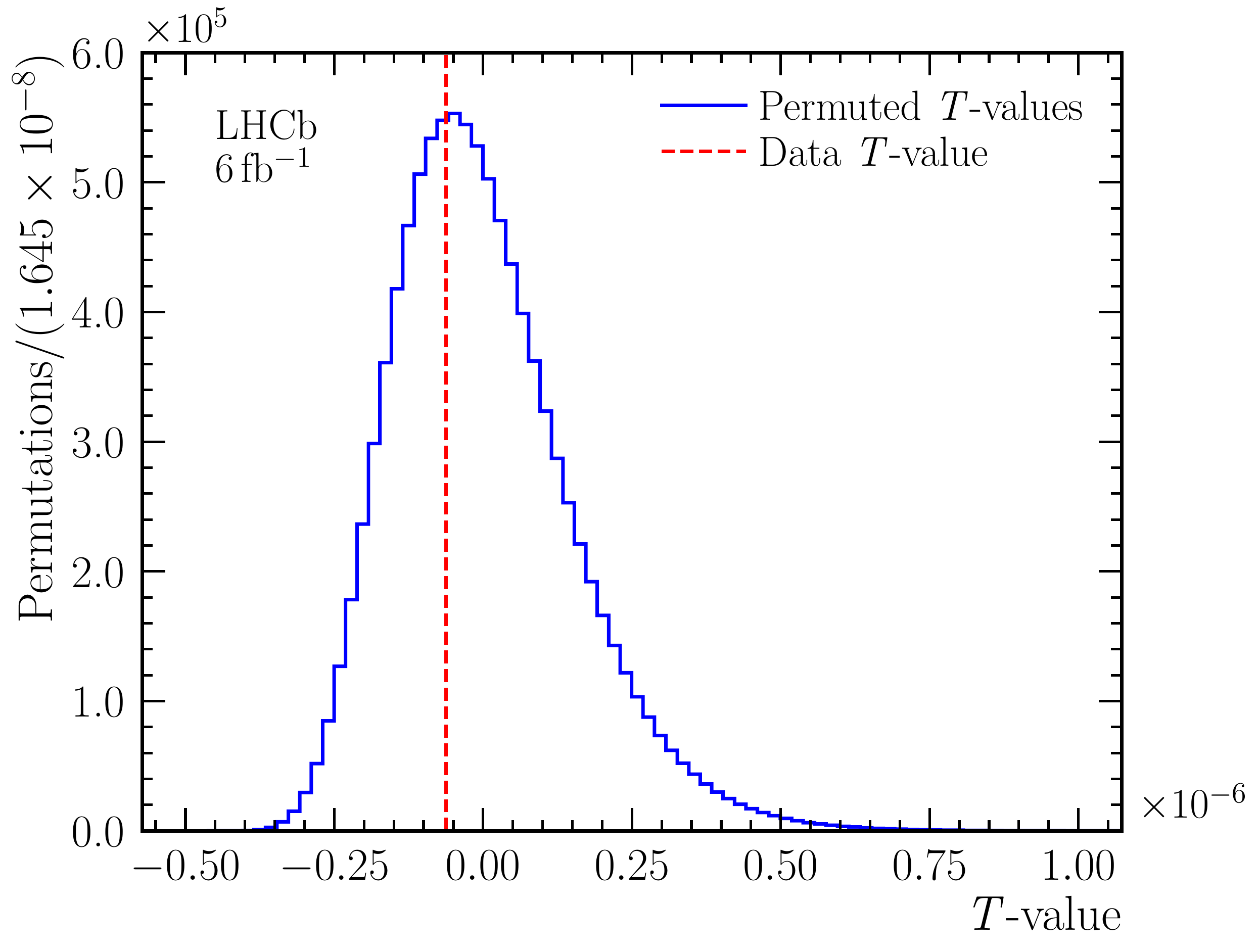}
\caption{\label{fig:finalpvalue} Distribution of $T$-values obtained by running the energy test over the final signal $\Dz \to \pi^-\pi^+\pi^0$ sample (red dashed line) superimposed on the corresponding distribution for the \CP-symmetry hypothesis, obtained from flavour-randomised permutations of the same data sample.}
\end{figure}

After comprehensive validation as discussed above, the energy test is performed in the $\Delta m$ signal region for the \mbox{$\Dz\to\pi^-\pi^+\pi^0$} channel. This yields a $p$-value of 0.62, giving no indication of any \CP violation in localised regions of the phase space. Figure~\ref{fig:finalpvalue} shows the corresponding $T$-value observed in data, superimposed on the expected distribution of $T$-values for the \CP-symmetry hypothesis. 

Similar cross-checks are performed as described for the control mode, including running the energy test separately for each data-taking year and hardware trigger category. All results are found to be self-consistent with no indication of systematic effects from nuisance asymmetries.

In summary, no evidence is found for \CP violation in localised regions of the phase space for the decay $\Dz \to \pi^-\pi^+\pi^0$. The analysis uses the complete data sample available in LHCb Run~2, collected during 2015--2018, containing approximately four times the signal yield of the previous energy test analysis of this channel based on data taken in Run~1 during 2012~\cite{LHCb:2014nnj}. 
Compared to the previous analysis, the present work benefits from a computationally more efficient energy test implementation, from a re-optimised candidate selection, and from a more comprehensive suite of data-driven cross-checks to confirm that potential nuisance asymmetries can be neglected for the current sample size.

\section*{Acknowledgements}
%
%
\noindent We express our gratitude to our colleagues in the CERN
accelerator departments for the excellent performance of the LHC. We
thank the technical and administrative staff at the LHCb
institutes.
We acknowledge support from CERN and from the national agencies:
CAPES, CNPq, FAPERJ and FINEP (Brazil); 
MOST and NSFC (China); 
CNRS/IN2P3 (France); 
BMBF, DFG and MPG (Germany); 
INFN (Italy); 
NWO (Netherlands); 
MNiSW and NCN (Poland); 
MEN/IFA (Romania); 
MICINN (Spain); 
SNSF and SER (Switzerland); 
NASU (Ukraine); 
STFC (United Kingdom); 
DOE NP and NSF (USA).
We acknowledge the computing resources that are provided by CERN, IN2P3
(France), KIT and DESY (Germany), INFN (Italy), SURF (Netherlands),
PIC (Spain), GridPP (United Kingdom), 
CSCS (Switzerland), IFIN-HH (Romania), CBPF (Brazil),
Polish WLCG  (Poland) and NERSC (USA).
We are indebted to the communities behind the multiple open-source
software packages on which we depend.
Individual groups or members have received support from
ARC and ARDC (Australia);
Minciencias (Colombia);
AvH Foundation (Germany);
EPLANET, Marie Sk\l{}odowska-Curie Actions, ERC and NextGenerationEU (European Union);
A*MIDEX, ANR, IPhU and Labex P2IO, and R\'{e}gion Auvergne-Rh\^{o}ne-Alpes (France);
Key Research Program of Frontier Sciences of CAS, CAS PIFI, CAS CCEPP, 
Fundamental Research Funds for the Central Universities, 
and Sci. \& Tech. Program of Guangzhou (China);
GVA, XuntaGal, GENCAT, Inditex, InTalent and Prog.~Atracci\'on Talento, CM (Spain);
SRC (Sweden);
the Leverhulme Trust, the Royal Society
 and UKRI (United Kingdom).

\addcontentsline{toc}{section}{References}
\bibliographystyle{LHCb}
\bibliography{main,standard,LHCb-PAPER,LHCb-CONF,LHCb-DP,LHCb-TDR}

\newpage
\centerline
{\large\bf LHCb collaboration}
\begin
{flushleft}
\small
R.~Aaij$^{32}$\lhcborcid{0000-0003-0533-1952},
A.S.W.~Abdelmotteleb$^{51}$\lhcborcid{0000-0001-7905-0542},
C.~Abellan~Beteta$^{45}$,
F.~Abudin{\'e}n$^{51}$\lhcborcid{0000-0002-6737-3528},
T.~Ackernley$^{55}$\lhcborcid{0000-0002-5951-3498},
B.~Adeva$^{41}$\lhcborcid{0000-0001-9756-3712},
M.~Adinolfi$^{49}$\lhcborcid{0000-0002-1326-1264},
P.~Adlarson$^{77}$\lhcborcid{0000-0001-6280-3851},
H.~Afsharnia$^{9}$,
C.~Agapopoulou$^{43}$\lhcborcid{0000-0002-2368-0147},
C.A.~Aidala$^{78}$\lhcborcid{0000-0001-9540-4988}
% ...

{\footnotesize \it

% ...
$^{9}$Universit{\'e} Clermont Auvergne, CNRS/IN2P3, LPC, Clermont-Ferrand, France\\
% ...
$^{32}$Nikhef National Institute for Subatomic Physics, Amsterdam, Netherlands\\
% ...
$^{41}$Instituto Galego de F{\'\i}sica de Altas Enerx{\'\i}as (IGFAE), Universidade de Santiago de Compostela, Santiago de Compostela, Spain\\
% ...
$^{43}$European Organization for Nuclear Research (CERN), Geneva, Switzerland\\
% ...
$^{45}$Physik-Institut, Universit{\"a}t Z{\"u}rich, Z{\"u}rich, Switzerland\\
% ...
$^{49}$H.H. Wills Physics Laboratory, University of Bristol, Bristol, United Kingdom\\
% ...
$^{51}$Department of Physics, University of Warwick, Coventry, United Kingdom\\
% ...
$^{54}$School of Physics and Astronomy, University of Glasgow, Glasgow, United Kingdom\\
$^{55}$Oliver Lodge Laboratory, University of Liverpool, Liverpool, United Kingdom\\
% ...
$^{63}$Syracuse University, Syracuse, NY, United States\\
% ...
$^{77}$Department of Physics and Astronomy, Uppsala University, Uppsala, Sweden, associated to $^{54}$\\
$^{78}$University of Michigan, Ann Arbor, MI, United States, associated to $^{63}$\\
\bigskip
% ...
}
\end{flushleft}

\end{document}